\documentclass[aps,prl,twocolumn,superscriptaddress,showpacs,floatfix,amsmath,amssymb]{revtex4-1}
\usepackage{graphicx}
\usepackage{dcolumn}
\usepackage{bm}
\usepackage{mathrsfs}

\begin{document}

\title{Nonlinear superposition of direct and inverse cascades in two-dimensional turbulence 
forced at large and small scales}

\author{Massimo Cencini} \affiliation{Istituto dei Sistemi Complessi,
  Consiglio Nazionale delle Ricerche, Via dei Taurini 19, I-00185
  Rome, Italy}

\author{Paolo Muratore-Ginanneschi} \affiliation{Department of
  Mathematics and Statistics, University of Helsinki, P\,B 68 Helsinki
  00014, Finland}

\author{Angelo Vulpiani} \affiliation{Dipartimento di Fisica,
  University ``Sapienza'', Piazzale A. Moro 2, 00185 Rome Italy}

\begin{abstract}
  We inquire about the properties of $2d$ Navier-Stokes turbulence
  simultaneously forced at small and large scales. The background
  motivation comes by observational results on atmospheric
  turbulence. We show that the velocity field is amenable to the sum
  of two auxiliary velocity fields forced at large and small scale and
  exhibiting a direct-enstrophy and an inverse-energy cascade,
  respectively. Remarkably, the two auxiliary fields reconcile
  universal properties of fluxes with positive statistical correlation
  in the inertial range.
\end{abstract}


\maketitle

Turbulence represents a tantalizing nonequilibrium system
characterized by cascade processes which, as typical in statistical
physics, strongly depends on space dimensionality. In $d\!\!=\!\!3$,
kinetic energy, injected at large scales, goes toward smaller ones
with positive and constant flux ($\approx\!$ energy-injection rate
$\epsilon$), until is dissipated by molecular
diffusion~\cite{Frisch1995}. Between the injection and dissipative
scales, the energy spectrum behaves as a power-law $E(k) \!\approx\!
\epsilon^{2/3} k^{-{5}/{3}}\!\!$. In $d\!=\!\!2$, ideal (inviscid and
unforced) fluids preserve both energy $\!\prec\! v^2\!\succ\!\!/2$ and
square vorticity (enstrophy) $\!\prec\!\omega^2\!\succ\!/2$ ($\omega\!
\!= \!\!\bm \nabla \!\times\!  \bm v$).  On this basis,
Kraichnan~\cite{Kraichnan1967} predicted that sustaining the flow at a
single scale $\ell_f\,(\sim\!k^{-1}_f)$, with energy (enstrophy)
injection rate $\epsilon$ ($\eta\!=\!\epsilon k_f^2$), generates a
double cascade of enstrophy downscale ($<\!\ell_f$) and of energy
upscale ($>\!\ell_f$). He also predicted two power laws for the energy
spectrum: $E(k)\!\approx\!  \eta^{2/3} k^{-3}$ (but for
log-corrections \cite{Kraichnan1971}) in the direct enstrophy cascade
range; $E(k)\!\approx\!  \epsilon^{2/3} k^{-{5}/{3}}$ for the inverse
energy cascade.  The direct cascade, with a positive enstrophy flux
($\!\approx\!  \eta$), ends at the dissipative scale.  Whilst, in an
unbounded domain, the inverse cascade proceeds undisturbed, with a
negative energy flux ($\approx\!\!-\epsilon$), unless large-scale
friction stops it at a scale $\!\gg\! \ell_f$ \cite{Boffetta2000}.
For a recent numerical study of the dual cascade see \cite{Boffetta2010}.

In $3d$-layers, as the atmosphere, both $3d$ and $2d$-phenomenology
can be relevant depending on the aspect ratio, the injection and
observation scales \cite{Smith1996,Celani2010,Xia2011}. Aircraft
measurements \cite{Nastrom1984,Lindborg1999} of atmospheric-winds
revealed that horizontal energy spectra at the troposphere end (at
$\approx\!  10$Km altitude) display two power-laws: $\!E(k)\!\propto\!
k^{-{5}/{3}}\!$ at wave-numbers in the mesoscales ($\approx\!\!
10\!-\!500$Km); $E(k)\!\propto\! k^{-3}$ at synoptic scales
($\approx\!\!  500\!-\!3,000$Km). Though, $2d$ phenomenology
should dominate at scales larger than the troposphere thickness
\cite{VallisBook}, measured spectra display the steeper and shallower
power-laws in reverse order with respect to Kraichnan's scenario.  To
complicate the picture, the energy flux seems to be positive at
$10\!-\!100$Km \cite{Cho2001}, suggesting a $3d$-like direct energy
cascade, though the involved scales may be too large.

In the $2d$-framework, on which we focus here, several explanations
for the observed spectra have been proposed.  Interpreting the
synoptic $-3$ spectrum as an enstrophy cascade, forced by
instabilities of the horizontal motion at the planetary scale
($\!\sim\!\! 10^4$Km) \cite{VallisBook}, the $-5/3$ mesoscales
spectrum may result: from a $2d$-inverse energy cascade forced by
convection driven by thermal gradients in the troposphere
\cite{Gage1979,FolkmarLarsen1982,Lilly1989}; or, less likely
\cite{Gage1986,Lindborg1999}, from gravity waves
\cite{Dewan1979,Vanzandt1982}.  Interpreting the $-5/3$ spectrum as a
convection-driven $2d$-inverse energy cascade, the steeper synoptical
spectrum may result from large-scale coherent structures due to forcing
\cite{Xia2008} or to energy condensation at the planetary scale
\cite{Smith1994,Xia2008}. Interestingly, such coherent motions
may mask the 
%underlying 
inverse cascade inducing a positive  energy-flux
\cite{Xia2008}, which could explain
observations \cite{Cho2001}.  Other proposed mechanisms (not discussed
here) consider $3d$ or quasi-$2d$ scenarios accounting for
stratification and other effects
\cite{Lindborg2006,Kitamura2006,Tulloch2006}.

In this Letter we focus on $2d$ turbulence forced at two, well
separated, scales with the aim of understanding the interplay of
oppositely directed cascades in the same range of scales.  We thus
consider the $2d$ incompressible Navier-Stokes equation, which for the
vorticity $\omega$ reads
\begin{equation}
\partial_t \omega \!+\!\bm v \cdot \bm \nabla \omega\!=
\!-\nu_p (-\Delta)^p \omega \!-\!\alpha_q (-\Delta)^{-q}\omega \!+\!f_{_{L}} \!+\!f_\ell \,,  
\label{eq:ns0} 
\end{equation}
$\bm v\!\!=\!\!\bm \nabla^\perp \psi\!\!=\!(\partial_y
\psi,\!-\partial_x \psi)$ is the velocity, and $\psi$ the stream
function ($\Delta \psi\!\!=\!\!-\omega$). The hyperviscous
(hypofriction) term removes enstrophy (energy) at small (large) scales
generalizing standard dissipation $p\!=\!1$ (Ekman friction
$q\!=\!0$).  In direct numerical simulations (DNS), such
generalizations provide for extended inertial ranges. We use $p\!=\!8$
and $q\!=\!1$; tests with different values have been also performed.
The forcings $f_{_L}$ and $f_\ell$ act at separate scales $L\!\!\gg
\!\!\ell$, injecting energy at scale $\ell$ and enstrophy at scale $L$,
at independent rates $\epsilon$ and $\eta$, respectively. We used two
independent random, zero-mean Gaussian processes restricted to a
narrow band in Fourier space $k\in [k_{i 1},k_{i 2}]$ (with
$i\!=\!\ell,L$) centered at $k_L\!\sim\! 1/L$ and $k_\ell\! \sim
\!1/\ell$, with correlation $\prec\!\! \hat{f}_i(\bm k,t)\hat{f}_j(\bm
k',t)\!\!\succ\!=\! F_i \delta_{ij}\delta(t\!-\!t')\delta(\bm
k\!+\!\bm k') \Theta(k\!-\!k_{i 1}) \Theta(k_{i 2}\!-\!k)$, $\Theta$
being the Heaviside step function. In the sequel, we ignore the
behavior at scales larger (smaller) than $L$ ($\ell$) as damped by
hypofriction (dissipation).

\begin{figure}[t!]
\centering
\includegraphics[width=0.43\textwidth]{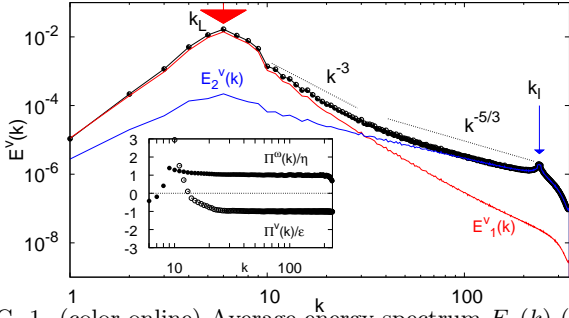}
\vspace{-0.7truecm}
\caption{(color online) Average energy spectrum $E_v(k)$
  (symbols). $E^{v}_{1,2}(k)$ labels the energy spectra associated with
  the auxiliary velocities $\bm v_{1,2}$ and suggests a direct
  (inverse) cascade for $\bm v_1$ ($\bm v_2$), see text.  Arrows mark
  the forcing bands at $k_L\!\approx\! 7$ and $k_\ell\!\approx\!
  240$. Dotted lines show the slopes $-3$ and $-5/3$. Inset: energy
  $\Pi^{v}(k)$ and enstrophy $\Pi^{\omega}(k)$ fluxes normalized by
  their average values ($\epsilon$ and $\eta$, resp.).  DNS of
  Eqs.~(\ref{eq:ns0}-\ref{eq:ns2}) were done with a standard
  $2/3$-dealiased, pseudospectal method with $1024^2$ collocation
  points.
\label{fig:1}}
\end{figure}
In Ref.~\cite{Gage1986}, to explain the observed universality of
atmospheric spectra, it was conjectured that the two sources may not
be independent. However, large and small scale excitations originate
from different physical processes \cite{VallisBook} characterized by
separate timescales (small-scale convection being faster), 
bearing their independence.
%makes a compelling case for assuming them as essentially independent. 
Within the independent-source model, spectral universality can be
ascribed to that of the inverse-cascade \cite{Boffetta2000}. In
\cite{FolkmarLarsen1982,Gage1986} it was also hypothesized that
oppositely directed cascades could not coexist 
without a sink between the forcing scales.
%unless a sink between the forcing scales was present.  
Lilly \cite{Lilly1989}, using closure
theories, showed that there is no need of such sink. Maltrud and
Vallis \cite{Maltrud1991} made, as far as we know, the 
%first and
unique numerical study of Eq.~(\ref{eq:ns0}), providing evidence of
two overlapping cascades.  This is confirmed by Fig.~\ref{fig:1} which
shows that the energy spectrum $E_v(k)$ displays the basic features of
the atmospheric one. Moreover, enstrophy $\Pi^{\omega}(k)$ and energy
$\Pi^{v}(k)$ fluxes are constant with opposite signs
($\Pi^{\omega}(k)>0$ and $\Pi^{v}(k)<0$) meaning that direct enstrophy
and inverse energy cascades superimpose, apparently undisturbed, in
the same range $[k_L:k_\ell]$.  We show below that, remarkably, the
superposition is realized maintaining non-trivial correlations between
the degrees of freedom associated with the two cascades.

To scrutinize this superposition we propose a decomposition able to
disentangle the two cascades,  by evolving in parallel with
Eq.~(\ref{eq:ns0}) two auxiliary equations
\begin{eqnarray}
&&\strut{\hspace{-.5truecm}}\partial_t \omega_{_L} +\bm v \cdot \bm \nabla \omega_{_L}\!\! = -\nu_p (-\Delta)^p \omega_{_L}\!\! -\alpha_q (-\Delta)^{-q} \omega_{_L}\!\! +f_{_{L}}  \label{eq:ns1} \\ 
&&\strut{\hspace{-.5truecm}}\partial_t \omega_\ell\, +\bm v \cdot \bm \nabla \omega_\ell\! = -\nu_p (-\Delta)^p \omega_\ell -\alpha_q  (-\Delta)^{-q}\omega_\ell\! +f_\ell  \label{eq:ns2} \,,
\end{eqnarray}
where $\boldsymbol{v}$ is the same as in Eq.~(\ref{eq:ns0}). Since
$f_{L}$ and $f_{\ell}$ in (\ref{eq:ns1}-\ref{eq:ns2}) are the same
realizations of the forcings in (\ref{eq:ns0}), $\omega_{{_L},\ell}$
are two ``active'' pseudo-scalar fields \cite{Celani2002},
such that $\omega\!=\!\omega_{_L}\!+\!\omega_\ell$ and $\bm v\!=\!\bm
v_{_L}\!+\!\bm v_\ell$ \cite{nota1},
with $\bm v_{{_L},\ell}\!=\!\bm
\nabla^{\perp}\psi_{{_L},\ell}$
($\Delta\psi_{{_L},\ell}=-\omega_{_{L},\ell}$) incompressible. In the
sequel, we adopt the notation $\bm v\!\equiv\!  \bm v_0$, $\omega
\!\equiv\! \omega_0$, $\bm v_{{_L},\ell} \!\equiv\! \bm v_{1,2}$ and
$\omega_{{_L},\ell}\!\equiv\!\omega_{1,2}$.

\begin{figure}[t!]
\centering
\includegraphics[width=0.48\textwidth]{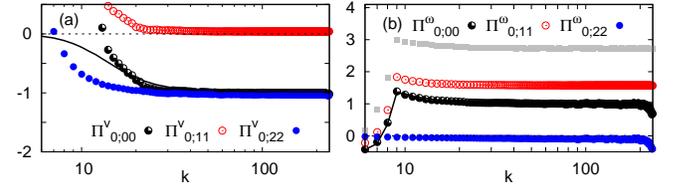} 
\vspace{-0.7truecm}
\caption{(color online) Dynamical fluxes 
 of (a) energy  $\Pi^{v}_{0;jj}(k)$ and (b) enstrophy $\Pi^{\omega}_{0;jj}(k)$ (defined in \cite{nota2}). 
Grey symbols in (b) refer to passive scalar flux $\Pi^\theta$, see text (Cfr. Eq.(\ref{eq:passive})). 
The solid curves show energy and enstrophy fluxes obtained by
  integrating (\ref{eq:ns1}) with $f_{L}=0$ (a) or
  $f_{\ell}=0$ (b), their superposition to $\Pi^v_{0;00}$ and $\Pi^{\omega}_{0;00}$, resp.
 further demonstrates the cascades overlap when both forcings are present.
Notice that $\Pi^v_{0;22}\approx \Pi^v_{0;00}<0$ and $\Pi^v_{0;11}\approx 0$ making evident that
$\bm v_2$ are the degrees of freedom associated with the inverse cascade.
Similar considerations apply to vorticity, though 
$\Pi^\omega_{0;11}>\Pi^\omega_{0;00}$, which  will be discussed later in relation to
  Fig.~\ref{fig:4}.
  \label{fig:2}}
\end{figure}

\begin{figure}[t!]
\centering
\includegraphics[width=0.48\textwidth]{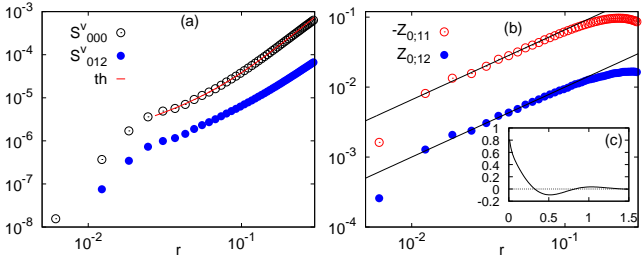}
\vspace{-0.7truecm}
\caption{(color online) $3^{rd}$ structure functions: (a) for
  velocity $S_{000}^{L}(r)$ compared with the prediction~(\ref{pred})
  (solid red line) and $S_{012}^{L}(r)$; (b) for vorticity
  $-Z_{0;11}(r)\!\equiv\!
  -\boldsymbol{r}\cdot\mathcal{Z}_{0;11}(\boldsymbol{r})/r$ and
  $Z_{0;12}(r)$, the black lines show linear behavior in
  $r$. The negative linear behavior of $Z_{0;11}(r)$ links to the direct cascade of $\prec\!\omega_1^2\!\succ$;
  (c) $\mathcal{C}^{\omega}_{12}(r)=\prec\!\omega_{1}(0)\omega_{2}(\boldsymbol{r})\!\succ$ vs $r$.
%Notice that 
The behaviors of $Z_{0;12}(r)$ (which should vanish in the ideal limit) and 
of $\mathcal{C}^{\omega}_{12}(r)$ are evidences of  the sensitivity of  directly cascading degrees 
of freedom to the hypofriction, see text (Cfr. Eq.~(\ref{nonlocal})). 
 \label{fig:3}}
\end{figure}

\begin{figure}[t!]
\centering
\includegraphics[width=0.48\textwidth]{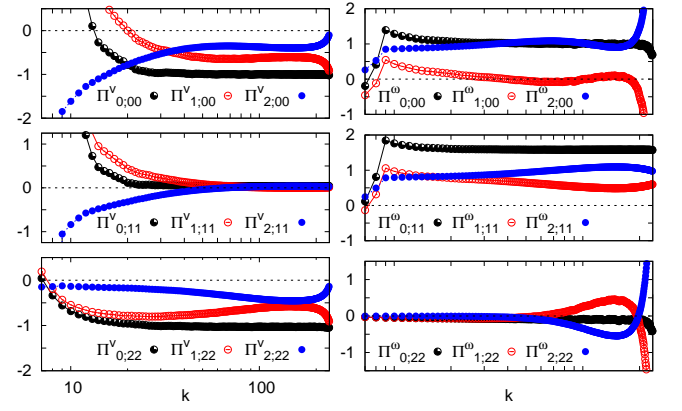}
\vspace{-0.7truecm}
\caption{(color online) Fluxes of energy and enstrophy defined in \cite{nota2}: 
(Left) energy fluxes $\Pi^{v}_{i;jj}$
  of $\bm v_j$ ($j=0,1,2$ from top to bottom) due to the transport
  by $\bm v_i$ ($i=0$ black semifilled circles, $i=1$ red empty
  circles, $i=2$ blue filled circles); (Right) Enstrophy fluxes
  $\Pi^{\omega}_{i;jj}$: panels, symbols and colors follow the same
  convention of Left panel.   \label{fig:4}}
\end{figure}

The behavior of the energy spectra (Fig.~\ref{fig:1}) and the fluxes
(Fig.~\ref{fig:2}) associated with the auxiliary fields suggest the
identification of the components $\boldsymbol{v}_{1,2}$ as the
carriers of the degrees of freedom mainly associated with, respectively,
the direct and the inverse cascade. This observation can be
substantiated by means of the K\'arm\'an-Howarth-Monin (KHM) equation
\cite{Frisch1995} $(\partial_t +\alpha_q (-\Delta)^{-q})C_{jk}(\bm
r,t)+\mathcal{E}_{jk}(\bm r,t)-F_{jk}(\bm r,t)=
\frac{1}{2}\partial_{\alpha} \mathcal{S}_{0jk}^{\alpha\beta\beta}(\bm
r,t)$ (as customary, we assumed translation, rotation and parity
invariance). The KHM equation links the $3^{rd}$ order
structure-tensor,
\begin{equation}
\mathcal{S}_{ijk}^{\alpha_{i}\alpha_{j}\alpha_{k}}(\boldsymbol{r})=
\prec\!\! \delta v_{i}^{\alpha_{i}}(\boldsymbol{r})\delta
v_{j}^{\alpha_{j}}(\boldsymbol{r}) \delta
v_{k}^{\alpha_{k}}(\boldsymbol{r})\!\! \succ 
\end{equation}
 ($ \delta
\boldsymbol{v}_{i}(\boldsymbol{r})\!\equiv\!\boldsymbol{v}_{i}(\boldsymbol{r},t)-\boldsymbol{v}_{i}(\boldsymbol{0},t)$),
to the correlation functions of velocities ($C_{jk}(\bm
r,t)\!=\prec\!\!\bm v_j(\bm r,t)\cdot \bm v_k(\bm 0,t)\!\!\succ$) and
forcings ($F_{jk}(\bm r)\!=\prec\!\! f_j(\bm r,t)f_k(\bm 0,t)\!\!\succ$) and
to the dissipative terms $\mathcal{E}_{jk}(\bm r,t)=2\nu_p \prec\!
\nabla_\alpha^p  v^\beta_j(\bm r,t) \nabla_\alpha^p v^\beta_k( \bm 0,t)\!\succ$.  Kraichnan's theory is equivalent to the
following three hypotheses \cite{Be99}: existence of steady state for
Galilean invariant statistical indicators; smoothness at finite
viscosity; absence of velocity dissipative anomaly.  
%Under these
%hypotheses, in the range $ \ell\ll r \ll L$, asymptotic expansions as
%in \cite{Be99} lead to the general form\
Using these hypotheses, a careful analysis along the lines of \cite{Be99,Lindborg1999,Mazzino2007} 
justifies in the range $ \ell\ll r \ll L$, the expansion
\begin{eqnarray}
\label{3tensor}
\lefteqn{\hspace{-1.0cm}
\mathcal{S}_{ijk}^{\alpha_{i}\alpha_{j}\alpha_{k}}(\boldsymbol{r})=
{\mbox {$\sum_{n=0,1}$}} r^{2\,n}
P_{ijk}\left\{ r^{\alpha_{i}}A_{\left\{i[jk]:n\right\}}\delta^{\alpha_{j}\alpha_{k}}\right\}
}
\nonumber\\ &&
-\,{\scriptstyle \frac{2}{3}}\,P_{ijk}\left\{A_{\left\{i[jk]:n\right\}}\right\}
\,r^{\alpha_{i}}r^{\alpha_{j}}r^{\alpha_{k}}+o( r^{3})\,,
\end{eqnarray}
with shorthand notation for summation over cyclic permutations,
$P_{ijk}\left\{O_{ijk}\right\}\!\equiv\!
O_{ijk}\!+\!O_{jki}\!+\!O_{kij}$ for any $O$.  
%Eq.~(\ref{3tensor}) strictly holds for $i=0$ 
%and,
%in the ideal limit of
%infinite volume at vanishing hypofriction $(\alpha_{q}=0)$,
%the constants are imposed by dynamics-grounded contraints 
Eq.~(\ref{3tensor}) holds true strictly for $i=0$
in the ideal limit of
infinite volume at vanishing hypofriction $(\alpha_{q}=0)$,
with, furthermore, the explicit prediction for the coefficients:
\begin{eqnarray}
\label{inverse}
\hspace{-0.2cm}
&&\strut{\hspace{-0.4truecm}}A_{\left\{0[jk]:0\right\}}\hspace{-0.06cm}
+\hspace{-0.06cm}P_{0jk}\left\{A_{\left\{0[jk]:0\right\}}\right\}
\hspace{-0.06cm}=\hspace{-0.06cm}
2\,\epsilon\,(\delta_{j0}\hspace{-0.06cm}+\hspace{-0.06cm}\delta_{j2})
(\delta_{k0}\hspace{-0.06cm}+\hspace{-0.06cm}\delta_{k2})
\\
&&\strut{\hspace{-0.4truecm}}A_{\left\{0[jk]:1\right\}}\hspace{-0.06cm}
+\hspace{-0.06cm}{\scriptstyle \frac{1}{3}}P_{0jk}\left\{\!A_{\left\{0[jk]:1\right\}}\!\right\}
\hspace{-0.06cm}=\hspace{-0.06cm}{\scriptstyle \frac{1}{4}}\eta\,(\delta_{j0}\hspace{-0.06cm}+\hspace{-0.06cm}\delta_{j1})
(\delta_{k0}\hspace{-0.06cm}+\hspace{-0.06cm}\delta_{k1}). \label{direct}
\end{eqnarray}
Out of the ideal limit, the constants
$A_{\left\{i[jk]:n\right\}}=A_{\left\{i[kj]:n\right\}}$ depend on the
full statistics of the solution of Eqs.~(\ref{eq:ns0}-\ref{eq:ns2}).
Moreover, for $i=1,2$, no dynamical constraints can be imposed to fix
the constants $A_{\left\{i[jk]:\{0,1\}\right\}}$, however the
expansion (\ref{3tensor}) can still be justified using parity
invariance and the incompressibility of the fields $\bm v_i$.  The
quantities $\mathcal{S}_{0jk}$ are thus associated with dynamical
fluxes, while $\mathcal{S}_{1jk}$ and $\mathcal{S}_{2jk}$ only provide
statistical information, and will be dubbed ``statistical'' fluxes.
Eqs.~(\ref{3tensor}-\ref{direct}) recover the $3^{rd}$-order
longitudinal structure function derived in Ref.~\cite{Lindborg1999}:
\begin{equation}
{\strut\hspace{-0.3truecm}}S_{000}^{L}(r)\!\equiv\! r_{\alpha_{1}}r_{\alpha_{2}}r_{\alpha_{3}}
\mathcal{S}_{000}^{\alpha_{1}\alpha_{2}\alpha_{3}}(r)/r^{3}={\scriptstyle\frac{3}{2}}\epsilon\,
r +{\scriptstyle\frac{1}{8}}\eta\, r^3\,,
\label{pred}
\end{equation}
which is numerically very well satisfied
(Fig.~\ref{fig:3}a). Interestingly, other quantities deviate from the
ideal limit prediction (e.g $S_{012}^{{L}}(r)\!=\!0$ is not reproduced
by data, see Fig.~\ref{fig:3}a). As discussed below, similar issues
are present for $3^{rd}$-order quantities associated with $\omega_i$
($i=0,1,2$). The deviations from ideal-limit predictions should be
ascribed to the finiteness of the simulation domain, where the
presence of hypofriction stymies the derivation of neat relations such
as (\ref{direct}).  The $A_{\left\{\dots:1\right\}}$'s become coupled
to the infra-red component of the kinetic energy hinting at a more
non-local and less universal behavior of the enstrophy cascade. This
fact relates to non-trivial correlations existing between the degrees
of freedom associated with the two cascades.  The importance of such
correlations can be quantified by comparing the ``statistical'' fluxes
$\Pi^{v,\omega}_{1;jj}(k)$ and $\Pi^{v,\omega}_{2;jj}(k)$ (with
$j=0,1,2$), defined in \cite{nota2}, with the ``dynamical'' fluxes
$\Pi^{v,\omega}_{0;jj}$'s (for $j=0,1,2)$.
%, which are imposed by energy balance. 
As shown in Fig.~\ref{fig:4}, two phenomena stand out.

Whilst $\Pi^v_{0;00} \simeq \Pi^v_{0;22}<0$ (Fig.~\ref{fig:2}a)
validates the identification of $\bm v_2$ as the degrees of freedom
associated with the inverse cascade, the left panels of Fig.~\ref{fig:4}
show that $\Pi^v_{1;00}<\! 0$ ($\Pi^v_{1;22}<\! 0$) with intensity
comparable to that of $\Pi^v_{2;00}$ ($\Pi^v_{2;22}$) and thus
indicates that $\bm v_1$ contributes to the inverse cascade of the
total field $\bm v_0$.  This may appear, at first glance, surprising
in consideration of the observed absence of flux of kinetic energy of
$\boldsymbol{v}_{1}$ ($\Pi^v_{i;11} \simeq\! 0$ for any $i$).  The
non-intuitive and, possibly, non-universal behavior of the statistical
energy fluxes $\Pi^v_{i;jj}$ ($i\!=\!1,2$) is, however, a
consequence of their dependence on the full statistics.  Similar
considerations apply to enstrophy fluxes, with the
roles of $\bm v_1$ and $\bm v_2$ exchanged (Fig.~\ref{fig:4} right).

The second phenomenon appertains to the relative intensity of
enstrophy fluxes. The ideal-limit energy-balance predictions
(\ref{inverse}-\ref{direct}) suggest
$\Pi_{0;00}^{\omega}\simeq\Pi_{0;11}^{\omega}\,>\,0$ with
$\Pi_{0;22}^{\omega}\simeq0$.  The right panels of Fig.~\ref{fig:4}
confirm the latter prediction coherently with the interpretation of
$\boldsymbol{v}_{2}$ as the carrier of the inverse cascade degrees of
freedom. They also show that the ``dynamical'' enstrophy flux
$\Pi_{0;11}^{\omega}$ is significantly enhanced with respect to
$\Pi_{0;00}^{\omega}$ (see also Fig.~\ref{fig:2}b).  The phenomenon
can be regarded as a consequence of the alleged fact that the main
contribution to $\boldsymbol{v}_{1}$ comes from the degrees of freedom
of $\boldsymbol{v}_{0}$ undergoing the direct cascade and of the
stronger sensitivity of this latter to non-local effects. In order to
substantiate the claim we inspected the equations governing the
correlation functions
$\mathcal{C}^{\omega}_{ij}(\boldsymbol{r})=\prec\!\!\omega_{i}(0)\omega_{j}(\boldsymbol{r})\!\!\succ$.
From the energy balance in the presence of hypofriction together with
the assumption of inverse cascading $\boldsymbol{v}_{2}$
(i.e. $\Pi^{\omega}_{0;22}\sim 0$ with
$\alpha_{q}\,(-\Delta)^{q}\mathcal{C}^{\omega}_{22}(\boldsymbol{0})=F_{\ell}=\prec\!f_\ell^2\!\succ$) we  derived that 
\begin{eqnarray}
\label{nonlocal}
&&{\strut\hspace{-0.3truecm}}
\Pi^{\omega}_{0;11}-\Pi^{\omega}_{0;00}=
{\scriptstyle \frac{1}{2}}\alpha_{q}\,(-\Delta)^{q}(\mathcal{C}^{\omega}_{00}-\mathcal{C}^{\omega}_{11})(\boldsymbol{0})
-F_{\ell}
\nonumber\\&&{\strut\hspace{0.3truecm}}
+O(k\,L)^{\gamma}\simeq \alpha_{q}\,(-\Delta)^{q}\mathcal{C}^{\omega}_{12}(\boldsymbol{0})
+O(k\,L)^{\gamma}
\end{eqnarray}
for
some $\gamma>\!0$, as enstrophy is bounded
\cite{Be99}. Eq.~(\ref{nonlocal}) evinces the non-local nature of the
enstrophy transfer. Its positive value implies that the correlation
$\mathcal{C}^{\omega}_{12}(r)$ must be also positive in the inertial
range, as confirmed in Fig.~\ref{fig:3}c.  In terms of the vorticity
structure functions
$\boldsymbol{\mathcal{Z}}_{i;jk}(\boldsymbol{r})\equiv\prec\!\delta\boldsymbol{v}_{i}(r)
\delta\omega_{j}(\boldsymbol{r})\delta\omega_{k}(\boldsymbol{r})\!\succ$,
Eq.~(\ref{nonlocal}) translates into
$Z_{0;12}(r)=\boldsymbol{r}\cdot\boldsymbol{\mathcal{Z}}_{0;12}(r)/r>0$
with linear dependence on $r$ (as shown in Fig.~\ref{fig:3}b), while it should be
$0$ in the ideal limit.
To further elucidate the enhancement of  $\Pi^\omega_{0;11}$ with
respect to $\Pi^\omega_{0;00}$, it is instructive to 
liken the ultra-violet behavior of the vorticity
$\omega_{1}$ to that of a passive scalar $\theta$ evolving in the same velocity
field, as in Ref.~\cite{Boffetta2002}: 
\begin{equation}
\partial_t \theta +\bm v \cdot \bm \nabla \theta\!\! = -\nu_p (-\Delta)^p \theta\!\! -\alpha_q (-\Delta)^{-q} \theta\!\! +f^\prime_{_{L}}  \label{eq:passive}
\end{equation}
Here, the large-scale forcing $f^\prime_{_L}$ has the same statistics
of that used in (\ref{eq:ns1}) but with independent realizations,
%ensuring that $\theta$ is a passive scalar
making $\theta$ a passive scalar
\cite{Celani2002}.  Flow incompressibility brings about a
scalar energy $\prec\! \theta^2\!\succ\!\!/2$ cascade to smaller scales,
and similarly to (\ref{nonlocal}), by energy balance, we have
\begin{eqnarray}
\label{scalar}
{\strut\hspace{-0.5truecm}}
\Pi^{\theta}\!-\!\Pi_{0;00}^{\omega}\simeq \alpha_{q}(-\partial_{\boldsymbol{x}}^{-2})^{q}(\mathcal{C}_{00}^{\omega}
-\mathcal{C}^{\theta})(\boldsymbol{0})/2+O(k\,L)^{\gamma^{\prime}}
\end{eqnarray}
for some positive $\gamma^\prime$, where $\mathcal{C}^{\theta}(\bm
r)=\prec\!\theta(\bm r,t)\theta(\bm 0,t)\!\succ$.  We expect the
difference in (\ref{scalar}) to be positive as the infra-red content
of $\mathcal{C}_{00}^{\omega}$ is ``fattened'' by the inverse energy
transfer at $k<k_L$. As shown in Fig.~\ref{fig:2}b, the comparison
between the passive scalar energy flux, $\Pi^\theta$, with the enstrophy flux,
$\Pi^\omega_{0;00}$ validates this prediction. The inference is that the
same phenomenon underlies the positive sign of (\ref{nonlocal}).

Summarizing, we showed that 2d-turbulence sustained by a large and a
small scale forcing gives rise in between the sources to an inertial
range where a direct and an inverse cascade co-exist and overlap.  We
also showed that there exists a natural decomposition of the
Navier--Stokes field compartmentalizing the degrees of freedom
associated with the direct and inverse cascade in two auxiliary velocity
fields, obtained considering a single large and small scale source,
respectively.  Although these auxiliary fields satisfy energy balance
relations as if they were independent, actually they are not and
exhibit non-trivial correlations pinpointed by the inspection of third
order statistics of ``statistical'' objects, evading the energy
balance relations.  In contrast to the settings used here, realistic
forcings in the atmosphere should be time-correlated, with the
large-scale excitation being slower than the small-scale one. Provided
the forcings are independent with separate spatial and temporal scale
the picture presented here should remain essentially
unaltered. However, a slow large-scale forcing may induce coherent
motions that, as argued in Ref.~\cite{Xia2008}, can change the sign of
the (total) energy flux and thus mask the inverse cascade process,
which may explain observations \cite{Cho2001}.

We conclude with a theoretical remark.  The decomposition in terms of
auxiliary velocity fields here proposed can be straightforwardly
generalized to $2d$ Navier-Stokes equations with an energy input
distributed over different scales. In this perspective, the cascade
overlap of the two sources model, here investigated, evinces the
physical mechanism for why Kraichnan theory applies also in the
presence of power-law sources and, consequently, for the inability of
renormalization group approach to correctly predict Navier-Stokes
energy spectra \cite{Mazzino2007}, even in what may seem a
priori a perturbative regime.

\begin{acknowledgments}
MC and AV acknowledge support from MIUR PRIN2009 ``Nonequilibrium
fluctuations: theory and applications''.
PMG acknowledges support from the Finnish Academy CoE ``Analysis and
Dynamics'' and from KITP (grant No. NSF PHY05-51164).
\end{acknowledgments}


\begin{thebibliography}{10}%
\makeatletter
\providecommand \@ifxundefined [1]{%
 \ifx #1\undefined \expandafter \@firstoftwo
 \else \expandafter \@secondoftwo
\fi
}%
\providecommand \@ifnum [1]{%
 \ifnum #1\expandafter \@firstoftwo
 \else \expandafter \@secondoftwo
\fi
}%
\providecommand \enquote [1]{``#1''}%
\providecommand \bibnamefont  [1]{#1}%
\providecommand \bibfnamefont [1]{#1}%
\providecommand \citenamefont [1]{#1}%
\providecommand\href[0]{\@sanitize\@href}%
\providecommand\@href[1]{\endgroup\@@startlink{#1}\endgroup\@@href}%
\providecommand\@@href[1]{#1\@@endlink}%
\providecommand \@sanitize [0]{\begingroup\catcode`\&12\catcode`\#12\relax}%
\@ifxundefined \pdfoutput {\@firstoftwo}{%
 \@ifnum{\z@=\pdfoutput}{\@firstoftwo}{\@secondoftwo}%
}{%
 \providecommand\@@startlink[1]{\leavevmode}%
 \providecommand\@@endlink[0]{}%
}{%
 \providecommand\@@startlink[1]{%
  \leavevmode
  \pdfstartlink
   attr{/Border[0 0 1 ]/H/I/C[0 1 1]}%
   user{/Subtype/Link/A<</Type/Action/S/URI/URI(#1)>>}%
  \relax
 }%
 \providecommand\@@endlink[0]{\pdfendlink}%
}%
\providecommand \url  [0]{\begingroup\@sanitize \@url }%
\providecommand \@url [1]{\endgroup\@href {#1}{\urlprefix}}%
\providecommand \urlprefix [0]{URL }%
\providecommand \Eprint[0]{\href }%
\@ifxundefined \urlstyle {%
  \providecommand \doi [1]{doi:\discretionary{}{}{}#1}%
}{%
  \providecommand \doi [0]{doi:\discretionary{}{}{}\begingroup
  \urlstyle{rm}\Url }%
}%
\providecommand \doibase [0]{http://dx.doi.org/}%
\providecommand \Doi[1]{\href{\doibase#1}}%
\providecommand \bibAnnote [3]{%
  \BibitemShut{#1}%
  \begin{quotation}\noindent
    \textsc{Key:}\ #2\\\textsc{Annotation:}\ #3%
  \end{quotation}%
}%
\providecommand \bibAnnoteFile [2]{%
  \IfFileExists{#2}{\bibAnnote {#1} {#2} {\input{#2}}}{}%
}%
\providecommand \typeout [0]{\immediate \write \m@ne }%
\providecommand \selectlanguage [0]{\@gobble}%
\providecommand \bibinfo [0]{\@secondoftwo}%
\providecommand \bibfield [0]{\@secondoftwo}%
\providecommand \translation [1]{[#1]}%
\providecommand \BibitemOpen[0]{}%
\providecommand \bibitemStop [0]{}%
\providecommand \bibitemNoStop [0]{.\EOS\space}%
\providecommand \EOS [0]{\spacefactor3000\relax}%
\providecommand \BibitemShut [1]{\csname bibitem#1\endcsname}%
%</preamble>
\bibitem{Frisch1995}%
  \BibitemOpen
  \bibfield{author}{%
  \bibinfo {author} {\bibfnamefont{U.}~\bibnamefont{Frisch}},\ }%
  \emph{\bibinfo {title} {Turbulence: the legacy of AN Kolmogorov}}\ (\bibinfo
  {publisher} {Cambridge University Press},\ \bibinfo {year} {1995})%
  \bibAnnoteFile{NoStop}{Frisch1995}%
\bibitem{Kraichnan1967}%
  \BibitemOpen
  \bibfield{author}{%
  \bibinfo {author} {\bibfnamefont{R.~H.}\ \bibnamefont{Kraichnan}},\ }%
  \bibfield{journal}{%
  \bibinfo {journal} {Phys. Fluids}\ }%
  \textbf{\bibinfo {volume} {10}},\ \bibinfo {pages} {1417} (\bibinfo {year}
  {1967})%
  \bibAnnoteFile{NoStop}{Kraichnan1967}%
\bibitem{Kraichnan1971}%
  \BibitemOpen
  \bibfield{author}{%
  \bibinfo {author} {\bibfnamefont{R.}~\bibnamefont{Kraichnan}},\ }%
  \bibfield{journal}{%
  \bibinfo {journal} {J. Fluid Mech.}\ }%
  \textbf{\bibinfo {volume} {47}},\ \bibinfo {pages} {525} (\bibinfo {year}
  {1971})%
  \bibAnnoteFile{NoStop}{Kraichnan1971}%
\bibitem{Boffetta2000}%
  \BibitemOpen
  \bibfield{author}{%
  \bibinfo {author} {\bibfnamefont{G.}~\bibnamefont{Boffetta}}, \bibinfo
  {author} {\bibfnamefont{A.}~\bibnamefont{Celani}},\ and\ \bibinfo {author}
  {\bibfnamefont{M.}~\bibnamefont{Vergassola}},\ }%
  \bibfield{journal}{%
  \Doi{10.1103/PhysRevE.61.R29}{\bibinfo {journal} {Phys. Rev. E}}\ }%
  \textbf{\bibinfo {volume} {61}},\ \bibinfo {pages} {29} (\bibinfo {year}
  {2000})%
  \bibAnnoteFile{NoStop}{Boffetta2000}%
\bibitem{Boffetta2010}%
  \BibitemOpen
  \bibfield{author}{%
  \bibinfo {author} {\bibfnamefont{G.}~\bibnamefont{Boffetta}}\ and\ \bibinfo
  {author} {\bibfnamefont{S.}~\bibnamefont{Musacchio}},\ }%
  \bibfield{journal}{%
  \Doi{10.1103/PhysRevE.82.016307}{\bibinfo {journal} {Phys. Rev. E}}\ }%
  \textbf{\bibinfo {volume} {82}},\ \bibinfo {pages} {016307} (\bibinfo {year}
  {2010})%
  \bibAnnoteFile{NoStop}{Boffetta2010}%
\bibitem{Smith1996}%
  \BibitemOpen
  \bibfield{author}{%
  \bibinfo {author} {\bibfnamefont{L.}~\bibnamefont{Smith}}, \bibinfo {author}
  {\bibfnamefont{J.}~\bibnamefont{Chasnov}},\ and\ \bibinfo {author}
  {\bibfnamefont{F.}~\bibnamefont{Waleffe}},\ }%
  \bibfield{journal}{%
  \Doi{10.1103/PhysRevLett.77.2467}{\bibinfo {journal} {Phys. Rev. Lett.}}\ }%
  \textbf{\bibinfo {volume} {77}},\ \bibinfo {pages} {2467} (\bibinfo {year}
  {1996})%
  \bibAnnoteFile{NoStop}{Smith1996}%
\bibitem{Celani2010}%
  \BibitemOpen
  \bibfield{author}{%
  \bibinfo {author} {\bibfnamefont{A.}~\bibnamefont{Celani}}, \bibinfo {author}
  {\bibfnamefont{S.}~\bibnamefont{Musacchio}},\ and\ \bibinfo {author}
  {\bibfnamefont{D.}~\bibnamefont{Vincenzi}},\ }%
  \bibfield{journal}{%
  \Doi{10.1103/PhysRevLett.104.184506}{\bibinfo {journal} {Phys. Rev. Lett.}}\
  }%
  \textbf{\bibinfo {volume} {104}},\ \bibinfo {pages} {184506} (\bibinfo {year}
  {2010})%
  \bibAnnoteFile{NoStop}{Celani2010}%
\bibitem{Xia2011}%
  \BibitemOpen
  \bibfield{author}{%
  \bibinfo {author} {\bibfnamefont{H.}~\bibnamefont{Xia}}, \bibinfo {author}
  {\bibfnamefont{D.}~\bibnamefont{Byrne}}, \bibinfo {author}
  {\bibfnamefont{G.}~\bibnamefont{Falkovich}},\ and\ \bibinfo {author}
  {\bibfnamefont{M.}~\bibnamefont{Shats}},\ }%
  \bibfield{journal}{%
  \Doi{10.1038/nphys1910}{\bibinfo {journal} {Nature Phys.}}\ }%
  \textbf{\bibinfo {volume} {7}},\ \bibinfo {pages} {321} (\bibinfo {year}
  {2011})%
  \bibAnnoteFile{NoStop}{Xia2011}%
\bibitem{Nastrom1984}%
  \BibitemOpen
  \bibfield{author}{%
  \bibinfo {author} {\bibfnamefont{G.}~\bibnamefont{Nastrom}}, \bibinfo
  {author} {\bibfnamefont{K.}~\bibnamefont{Gage}},\ and\ \bibinfo {author}
  {\bibfnamefont{W.}~\bibnamefont{Jasperson}},\ }%
  \bibfield{journal}{%
  \bibinfo {journal} {Nature}\ }%
  \textbf{\bibinfo {volume} {310}},\ \bibinfo {pages} {36} (\bibinfo {year}
  {1984})%
  \bibAnnoteFile{NoStop}{Nastrom1984}%
\bibitem{Lindborg1999}%
  \BibitemOpen
  \bibfield{author}{%
  \bibinfo {author} {\bibfnamefont{E.}~\bibnamefont{Lindborg}},\ }%
  \bibfield{journal}{%
  \Doi{10.1017/S0022112099004851}{\bibinfo {journal} {J. Fluid Mech.}}\ }%
  \textbf{\bibinfo {volume} {388}},\ \bibinfo {pages} {259} (\bibinfo {year}
  {1999})%
  \bibAnnoteFile{NoStop}{Lindborg1999}%
\bibitem{VallisBook}%
  \BibitemOpen
  \bibfield{author}{%
  \bibinfo {author} {\bibfnamefont{G.}~\bibnamefont{Vallis}},\ }%
  \emph{\bibinfo {title} {Atmospheric and oceanic fluid dynamics: fundamentals
  and large-scale circulation}}\ (\bibinfo {publisher} {Cambridge University
  Press},\ \bibinfo {year} {2006})%
  \bibAnnoteFile{NoStop}{VallisBook}%
\bibitem{Cho2001}%
  \BibitemOpen
  \bibfield{author}{%
  \bibinfo {author} {\bibfnamefont{J. Y. N.}~\bibnamefont{Cho}}\ and\ \bibinfo
  {author} {\bibfnamefont{E.}~\bibnamefont{Lindborg}},\ }%
  \bibfield{journal}{%
  \bibinfo {journal} {J. Geophys. Res}\ }%
  \textbf{\bibinfo {volume} {106}},\ \bibinfo {pages} {223} (\bibinfo {year}
  {2001})%
  \bibAnnoteFile{NoStop}{Cho2001}%
\bibitem{Gage1979}%
  \BibitemOpen
  \bibfield{author}{%
  \bibinfo {author} {\bibfnamefont{K.}~\bibnamefont{Gage}},\ }%
  \bibfield{journal}{%
  \bibinfo {journal} {J. Atmos. Sci.}\ }%
  \textbf{\bibinfo {volume} {36}},\ \bibinfo {pages} {1950} (\bibinfo {year}
  {1979})%
  \bibAnnoteFile{NoStop}{Gage1979}%
\bibitem{FolkmarLarsen1982}%
  \BibitemOpen
  \bibfield{author}{%
  \bibinfo {author} {\bibfnamefont{M.}~\bibnamefont{Folkmar~Larsen}}, \bibinfo
  {author} {\bibfnamefont{M.}~\bibnamefont{Kelley}},\ and\ \bibinfo {author}
  {\bibfnamefont{K.}~\bibnamefont{Gage}},\ }%
  \bibfield{journal}{%
  \bibinfo {journal} {J. Atmos. Sci.}\ }%
  \textbf{\bibinfo {volume} {39}},\ \bibinfo {pages} {1035} (\bibinfo {year}
  {1982})%
  \bibAnnoteFile{NoStop}{FolkmarLarsen1982}%
\bibitem{Lilly1989}%
  \BibitemOpen
  \bibfield{author}{%
  \bibinfo {author} {\bibfnamefont{D.}~\bibnamefont{Lilly}},\ }%
  \bibfield{journal}{%
  \bibinfo {journal} {J. Atmos. Sci.}\ }%
  \textbf{\bibinfo {volume} {46}},\ \bibinfo {pages} {2026} (\bibinfo {year}
  {1989})%
  \bibAnnoteFile{NoStop}{Lilly1989}%
\bibitem{Gage1986}%
  \BibitemOpen
  \bibfield{author}{%
  \bibinfo {author} {\bibfnamefont{K.}~\bibnamefont{Gage}}\ and\ \bibinfo
  {author} {\bibfnamefont{G.}~\bibnamefont{Nastrom}},\ }%
  \bibfield{journal}{%
  \bibinfo {journal} {J. Atmos. Sci.}\ }%
  \textbf{\bibinfo {volume} {43}},\ \bibinfo {pages} {729} (\bibinfo {year}
  {1986})%
  \bibAnnoteFile{NoStop}{Gage1986}%
\bibitem{Dewan1979}%
  \BibitemOpen
  \bibfield{author}{%
  \bibinfo {author} {\bibfnamefont{E.}~\bibnamefont{Dewan}},\ }%
  \bibfield{journal}{%
  \bibinfo {journal} {Science}\ }%
  \textbf{\bibinfo {volume} {204}},\ \bibinfo {pages} {832} (\bibinfo {year}
  {1979})%
  \bibAnnoteFile{NoStop}{Dewan1979}%
\bibitem{Vanzandt1982}%
  \BibitemOpen
  \bibfield{author}{%
  \bibinfo {author} {\bibfnamefont{T.}~\bibnamefont{VanZandt}},\ }%
  \bibfield{journal}{%
  \bibinfo {journal} {Geophys. Res. Lett.}\ }%
  \textbf{\bibinfo {volume} {9}},\ \bibinfo {pages} {575} (\bibinfo {year}
  {1982})%
  \bibAnnoteFile{NoStop}{Vanzandt1982}%
\bibitem{Xia2008}%
  \BibitemOpen
  \bibfield{author}{%
  \bibinfo {author} {\bibfnamefont{H.}~\bibnamefont{Xia}}, \bibinfo {author}
  {\bibfnamefont{H.}~\bibnamefont{Punzmann}}, \bibinfo {author}
  {\bibfnamefont{G.}~\bibnamefont{Falkovich}},\ and\ \bibinfo {author}
  {\bibfnamefont{M.}~\bibnamefont{Shats}},\ }%
  \bibfield{journal}{%
  \Doi{10.1103/PhysRevLett.101.194504}{\bibinfo {journal} {Phys. Rev. Lett.}}\
  }%
  \textbf{\bibinfo {volume} {101}},\ \bibinfo {pages} {194504} (\bibinfo {year}
  {2008})%
  \bibAnnoteFile{NoStop}{Xia2008}%
\bibitem{Smith1994}%
  \BibitemOpen
  \bibfield{author}{%
  \bibinfo {author} {\bibfnamefont{L.}~\bibnamefont{Smith}}\ and\ \bibinfo
  {author} {\bibfnamefont{V.}~\bibnamefont{Yakhot}},\ }%
  \bibfield{journal}{%
  \Doi{10.1017/S0022112094002065}{\bibinfo {journal} {J. Fluid Mech.}}\ }%
  \textbf{\bibinfo {volume} {274}},\ \bibinfo {pages} {115} (\bibinfo {year}
  {1994})%
  \bibAnnoteFile{NoStop}{Smith1994}%
\bibitem{Lindborg2006}%
  \BibitemOpen
  \bibfield{author}{%
  \bibinfo {author} {\bibfnamefont{E.}~\bibnamefont{Lindborg}},\ }%
  \bibfield{journal}{%
  \Doi{10.1017/S0022112005008128}{\bibinfo {journal} {J. Fluid Mech.}}\ }%
  \textbf{\bibinfo {volume} {550}},\ \bibinfo {pages} {207} (\bibinfo {year}
  {2006})%
  \bibAnnoteFile{NoStop}{Lindborg2006}%
\bibitem{Kitamura2006}%
  \BibitemOpen
  \bibfield{author}{%
  \bibinfo {author} {\bibfnamefont{Y.}~\bibnamefont{Kitamura}}\ and\ \bibinfo
  {author} {\bibfnamefont{Y.}~\bibnamefont{Matsuda}},\ }%
  \bibfield{journal}{%
  \bibinfo {journal} {Geophys. Res. Lett.}\ }%
  \textbf{\bibinfo {volume} {33}},\ \bibinfo {pages} {L05809} (\bibinfo {year}
  {2006})%
  \bibAnnoteFile{NoStop}{Kitamura2006}%
\bibitem{Tulloch2006}%
  \BibitemOpen
  \bibfield{author}{%
  \bibinfo {author} {\bibfnamefont{R.}~\bibnamefont{Tulloch}}\ and\ \bibinfo
  {author} {\bibfnamefont{K.}~\bibnamefont{Smith}},\ }%
  \bibfield{journal}{%
  \Doi{10.1073/pnas.0605494103}{\bibinfo {journal} {Proc. Natl. Acad. Sci.}}\
  }%
  \textbf{\bibinfo {volume} {103}},\ \bibinfo {pages} {14690} (\bibinfo {year}
  {2006})%
  \bibAnnoteFile{NoStop}{Tulloch2006}%
\bibitem{Maltrud1991}%
  \BibitemOpen
  \bibfield{author}{%
  \bibinfo {author} {\bibfnamefont{M.}~\bibnamefont{Maltrud}}\ and\ \bibinfo
  {author} {\bibfnamefont{G.}~\bibnamefont{Vallis}},\ }%
  \bibfield{journal}{%
  \bibinfo {journal} {J. Fluid Mech.}\ }%
  \textbf{\bibinfo {volume} {228}},\ \bibinfo {pages} {321} (\bibinfo {year}
  {1991})%
  \bibAnnoteFile{NoStop}{Maltrud1991}%
\bibitem{Celani2002}%
  \BibitemOpen
  \bibfield{author}{%
  \bibinfo {author} {\bibfnamefont{A.}~\bibnamefont{Celani}}, \bibinfo {author}
  {\bibfnamefont{M.}~\bibnamefont{Cencini}}, \bibinfo {author}
  {\bibfnamefont{A.}~\bibnamefont{Mazzino}},\ and\ \bibinfo {author}
  {\bibfnamefont{M.}~\bibnamefont{Vergassola}},\ }%
  \bibfield{journal}{%
  \Doi{10.1103/PhysRevLett.89.234502}{\bibinfo {journal} {Phys. Rev. Lett.}}\
  }%
  \textbf{\bibinfo {volume} {89}},\ \bibinfo {pages} {234502} (\bibinfo {year}
  {2002})\bibinfo {note} {; {N}ew J. Phys. \textbf{6}, 72 (2004).}%
  \bibAnnoteFile{Stop}{Celani2002}%
\bibitem{nota1}%
  \BibitemOpen
  \bibinfo {note} {In simulations it is enough to integrate Eqs.(\ref {eq:ns0})
  and (\ref {eq:ns1}) and obtain $\omega _\ell $ by subtraction.}%
  \bibAnnoteFile{Stop}{nota1}%
\bibitem{nota2}%
  \BibitemOpen
  \bibinfo {note} {$\Pi ^{v}_{i;jj}(k)=\DOTSI \intop \ilimits@ _{q>k} \protect
  \frac {d^{2}q}{(2\protect \tmspace +\thinmuskip {.1667em}\pi )^{2}}\DOTSI
  \intop \ilimits@ _{\protect \mathbb {R}^{2}}d\protect \bm {r}\protect
  \tmspace +\thinmuskip {.1667em}e^{-\imath \protect \bm {q}\cdot \protect \bm
  {r}}\protect \tmspace +\thinmuskip {.1667em}\nabla _{\alpha } \protect
  \mathcal {S}_{ijj}^{\alpha \beta \beta }(\protect \bm {r})$, and $\Pi
  ^{\omega }_{i;jj}(k)=\DOTSI \intop \ilimits@ _{q>k} \protect \frac
  {d^{2}q}{(2\protect \tmspace +\thinmuskip {.1667em}\pi )^{2}}\DOTSI \intop
  \ilimits@ _{\protect \mathbb {R}^{2}}d\protect \bm {r}x\protect \tmspace
  +\thinmuskip {.1667em}e^{-\imath \protect \bm {q}\cdot \protect \bm
  {r}}\protect \tmspace +\thinmuskip {.1667em}\nabla _{\alpha } (-\Delta
  )\protect \mathcal {S}_{ijj}^{\alpha \beta \beta }(\protect \bm {r})$, with
  Einstein convention on repeated Greek indexes. Note that, by construction,
  $\Pi ^{v,\omega }_{1;jj}\!+\!\Pi ^{v,\omega }_{2;jj}\!\!=\!\Pi ^{v,\omega }_{0;jj}$ for
  any $j$}%
  \bibAnnoteFile{NoStop}{nota2}%
\bibitem{Be99}%
  \BibitemOpen
  \bibfield{author}{%
  \bibinfo {author} {\bibfnamefont{D.}~\bibnamefont{Bernard}},\ }%
  \bibfield{journal}{%
  \bibinfo {journal} {Phys. Rev. E}\ }%
  \textbf{\bibinfo {volume} {60}},\ \bibinfo {pages} {6184} (\bibinfo {year}
  {1999})%
\bibinfo {note} {; {E}urophys. Lett. \textbf{50}, 333 (2000)}%
  \bibAnnoteFile{Stop}{Celani2002}%
  \bibAnnoteFile{NoStop}{Be99}%
\bibitem{Boffetta2002}%
  \BibitemOpen
  \bibfield{author}{%
  \bibinfo {author} {\bibfnamefont{G.}~\bibnamefont{Boffetta}}, \bibinfo
  {author} {\bibfnamefont{A.}~\bibnamefont{Celani}}, \bibinfo {author}
  {\bibfnamefont{S.}~\bibnamefont{Musacchio}},\ and\ \bibinfo {author}
  {\bibfnamefont{M.}~\bibnamefont{Vergassola}},\ }%
  \bibfield{journal}{%
  \Doi{10.1103/PhysRevE.66.026304}{\bibinfo {journal} {Phys. Rev. E}}\ }%
  \textbf{\bibinfo {volume} {66}},\ \bibinfo {pages} {026304} (\bibinfo {year}
  {2002})%
  \bibAnnoteFile{NoStop}{Boffetta2002}%
\bibitem{Mazzino2007}%
  \BibitemOpen
  \bibfield{author}{%
  \bibinfo {author} {\bibfnamefont{A.}~\bibnamefont{Mazzino}}, \bibinfo
  {author} {\bibfnamefont{P.}~\bibnamefont{Muratore-Ginanneschi}},\ and\
  \bibinfo {author} {\bibfnamefont{S.}~\bibnamefont{Musacchio}},\ }%
  \bibfield{journal}{%
  \Doi{10.1103/PhysRevLett.99.144502}{\bibinfo {journal} {Phys. Rev. Lett.}}\
  }%
  \textbf{\bibinfo {volume} {99}},\ \bibinfo {pages} {144502} (\bibinfo {year}
  {2007})\bibinfo {note} {; {J}STAT, P10012 (2009).}%
  \bibAnnoteFile{Stop}{Mazzino2007}%
\end{thebibliography}
\end{document}